# Timed Test Case Generation Using Labeled Prioritized Time Petri Nets

Noureddine ADJIR[1], Pierre de SAQUI SANNES[2], M. Kamel RAHMOUNI[3] and Abdelkader ADLA[3]

[1] LMMC, Department of Informatics and Mathematics, University of Moulay Tahar, Saida,
BP 138, Ennasr, 20002, Saida, Algeria
*adjir_nourd@yahoo.fr*

[2] CNRS ; LAAS ; 7 avenue du Colonel Roche, F-31077 Toulouse, France
Université de Toulouse ; UPS, INSA, INP, ISAE ; LAAS ; F-31077 Toulouse France
*pdss@isae.fr*

[3] Department of Informatics, Faculty of Science, University of Oran,
BP 1524, El M'naouar, 31000, Oran, Algeria
*{kamel_rahmouni, aekadla}@yahho.fr*

**Abstract**

Model-based testing of software and hardware systems uses behavioral and formal models of the systems. The paper presents a technique for model-based black-box conformance testing of real-time systems using Labeled Prioritized Time Petri Nets (LPrTPN). The Timed Input/Output Conformance (tioco) relation, which takes environment assumptions into account, serves as reference to decide of implementation correctness. Test suites are derived automatically from a LPrTPN made up of two concurrent sub-nets that respectively specify the system under test and its environment. The result is optimal in the sense that test cases have the shortest possible accumulated time to be executed. Test cases selection combines test purposes and structural coverage criteria associated with the model. A test purpose or a coverage criterion is specified in a SE-LTL formula. The TIme Petri Net Analyzer TINA has been extended to support concurrent composed subnets. Automatic generation of time-optimal test suites with the Tina toolbox combines the model checker selt and the path analyzer plan. *selt* outputs a sequence that satisfies the logic formula. *plan* computes the fastest execution of this sequence which will be transformed in a test cases suite.

## 1. Introduction

The embedded real-time industry is changing fast – systems have become larger, more complex, and more integrated. Real-Time systems interact with their surrounding environment and provide the latter with the expected outputs at the right time. In other words, the timely reaction is just as important as the kind of reaction. Such systems need to be tested in order to check their reliability before use. Testing real-time systems is even more challenging than testing untimed ones, because the tester must consider when to stimulate the system, when to expect responses to be issued, and how to assign verdicts to any timed event sequence it may observe and partly control. Further, the test cases must be executed in real-time, which means the test execution system itself becomes a real-time system.

Without automation and modeling tools, testing remains ad hoc, error prone, and very expensive both at the level of the test suit construction and at execution time. Clearly, real-time testing is almost impossible to achieve manually for real-size systems.

With the use of models in software/hardware design and development, model-based testing has received increasing attention from industry practitioners. It is a black-box approach in which common testing tasks such as test case generation and test result evaluation are based on a model of the system. Using the model to generate test cases and assign verdicts is cheaper and more effective than a completely manual approach.

The paper addresses model-based black-box conformance testing of real-time systems. It checks a System Under Test (SUT) against its specification. This is typically achieved in a controlled environment where the SUT is executed and stimulated with inputs according to a test specification, and the responses of the SUT are checked to conform to its specification.

The paper advocates for a type of conformance testing where test suites are derived from a formal model that specifies the expected behavior of the system to be tested. Precisely, the paper presents a technique for model-based black-box conformance testing of real-time systems based on Labelled Prioritized Time Petri Nets models (LPrTPN). The test specification is given as an LPrTPN made up of two concurrent subnets that respectively model the expected behaviour of the SUT and the latter's environment.

Optimizing test case generation requires selecting a limited set of test cases to be executed from a very large, may be of unbounded size, list of tests that cover all the executions of the SST. Practically, a huge number of test cases, generally infinitely, can be generated from even the simplest models. The addition of real-time complicates matters and is a source of explosion of system states and consequently of test cases. To guide the test cases selection, a test purpose or coverage criterions are often used. Test purposes and coverage approaches guarantee that test suites are derived systematically. Further, the approaches "coverage criteria" guarantee a certain level of reliability, quality, thoroughness and confidence.

In this paper, test cases can be generated in two different manners: on the one hand, by using manually formulated test purposes then encoded in the SE-LTL logic [16]; on the other hand, by using several kinds of coverage criterion expressed directly in SE-LTL such as statements, transitions, places, markings or states classes coverage of the LPrTPN model. In this context, the paper explains how to exploit and extend the toolbox TINA [3] to calculate the test cases and test suites. The proposed approach exploits the fact that TINA includes the tools *selt* and *plan*. *selt* is a State-Event LTL model checker. *plan* is a path analysis tool that computes a firing schedule over some given firing transition sequences, in particular, the fastest and the shortest schedules. The latest release of TINA supports automatic generation of time-optimal test suites for conformance testing i.e. test suites with optimal execution time. So, the particular schedules, computed by plan, associated to the witnesses sequences of a test purpose or a coverage criteria exhibited by *selt*, will be used to compute the time-optimal test cases and test suites from the SUT and the considering environment models. Especially, the required behaviour of the SUT is specified using a Deterministic Input Enabled and Output Urgent LPrTPN (DIOU-LPrTPN).

Time-optimal test suites are interesting for several reasons. First, reducing the total execution time of a test suite allows more behaviour to be tested in the (limited) time allocated to testing; we may thus expect tests to be more thorough. Secondly, it is generally desirable that regression testing can be executed as quickly as possible to improve the turnaround time between changes. Thirdly, it is essential for product instance testing that a thorough test can be performed without testing becoming the bottleneck, i.e., the test suite must be applied to all products coming of an assembly line. Finally, in the context of testing of real-time systems, we hypothesize that the fastest test case that drives the SUT to some state, also has a high likelihood of detecting errors, because this is a stressful situation for the SUT to handle.

The rest of the paper is organized as follows: section 2 surveys related work. In section 3, we define the test specification. Section 4 defines the syntax and the semantics of the LPrTPN. It also discusses test case generation based on the DIOU-LPrTPN model. Section 5 describes how to encode test purposes and coverage criteria in the SE-LTL logic. Section 6 concludes the paper.

## 2. Related work and motivations

Time Petri nets [30] are one among the important formal models widely used to specify and verify real-time systems. They are characterized by their expressive power of parallelism and concurrency, and the conciseness of the models. In addition, the efficient analysis methods proposed by [10] have contributed to their wide use. Adding priorities to TPN (PrTPN) increases their expressiveness [2] and [9]. Since we address the testing of reactive systems, we associate a label of an alphabet of actions with each transition (LPrTPN). A label is an input or an output or an internal action.

TPN have other important advantages that are not mentioned here due to lack of space. Despite of this, little work has been done on model-based testing from TPNs, the subject being essentially addressed for timed automata (TA) [4]. Model-based testing for TA has been discussed in [12], [13], [14], [17], [19], [20], [22], [23], [24], [26], [27], [28], [31], [32] and [33], just to mention a few. Further, most TA-based testing tools were developed more than five years ago (see, e.g., [18], [22] and [31]).

Algorithms for generating test suites following test purposes or a coverage criteria attempt to optimize test suites w.r.t. the number of test cases, the total length of the test suite, and the total time required to execute the test suite. In the paper, we are interested in the last two propositions. In this context, the main contributions of the paper are as follows: re-implement the toolbox TINA and add functionalities to support the composition of LPrTPN's, definition of a subclass of LPrTPN from which the schedules computed by the path analysis tool plan, in particular the fastest schedules (optimal in the total time) and the shortest paths (optimal in the total length), associated to the diagnostic sequences, exhibited by the State-Event LTL model-checker selt [16], will be exploited to compute the time-optimal (covering) test suites.

## 3. Test specification

Testing involves a system surrounded by an environment. It is almost impossible to test the system without making assumptions about its environment. An uncontrolled and possibly imaginary environment would indeed allow all possible interaction sequences. But, due to the lack of resources, it is not feasible to validate the system for all possible environments. Practically, the requirements and the assumptions of the environment need to be made explicit.

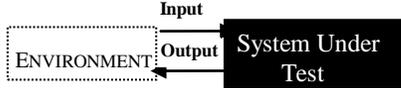

Figure 1.  The SUT and its environment ENV

We assume that the test specification, noted $\mathcal{M}=\mathcal{M}_{SUT} \| \mathcal{M}_E$, is given as an LPrTPN made up of two concurrent subnets. The first subnet models the expected behavior of the SUT, noted $\mathcal{M}_{SUT}$. The second subnet models the behavior of the environment; it is noted $\mathcal{M}_E$ (Fig. 2). The set of observable actions is partitioned into two subsets: input actions noted $A_{in}$ and output actions noted $A_{out}$. Inputs are the stimuli received by the system from the environment. Outputs are the actions sent by this system to its environment. They are not controllable and should be tested also with their deliverance dates. An input $a$ (resp. output $b$) is post fixed by $a?$ ($b!$). The system may perform internal actions which are invisible to the environment and thus to the tester. Internal or unobservable actions are denoted $\tau$. We have $A_\tau = A_{in} \cup A_{out} \cup \{\tau\}$.

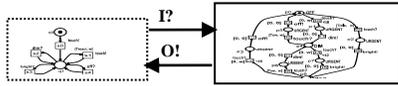

Figure 2. The test specification $\mathcal{M}$: The SUT model $\mathcal{M}_{SUT}$ and its environment model $\mathcal{M}_E$.

## 4. Environment and system modeling

### 4.1 Labelled Prioritized Time Petri Nets

Time Petri Nets (TPN) [30] extend Petri Nets with temporal intervals on transitions. PrTPN extend TPN with a priority relation on the transitions. Since we address the testing of reactive systems, we add an alphabet of actions $A_\tau$ and a labelling function that associates an action with each transition.

### 4.1.1 Notations

The sets $\mathbb{N}, \mathbb{Q}, \mathbb{Q}_{\geq 0}, \mathbb{R}, \mathbb{R}_{\geq 0}$ are respectively the sets of natural, rational, non-negative rational, real and non-negative real numbers. We consider the set $\mathbb{I}^+$ of non-empty real intervals $[a,b]$ with bounds $a,b \in \mathbb{Q}_{\geq 0}$. We consider both open and closed bounds, and also allow a right open infinite bound as in $[1,\infty[$. For $i \in \mathbb{I}^+$, $\downarrow i$ represents its lower bound, and $\uparrow i$ its superior bound (if it exists) or $\infty$. For any $\theta \in \mathbb{R}_{\geq 0}$, $i \dot{-} \theta$ represents the interval $\{x-\theta / x \geq \theta \wedge \theta \geq 0\}$. $A_S = A_{in} \times A_{out} \cup A_{out} \times A_{in}$ is the set of the couples of synchronizing actions and $A_{S\tau} = A_s \cup \{\tau\}$ is the set of all actions (internal and synchronizing actions).

### 4.1.2 Syntax

Formally, a LPrTPN over the alphabet $A_\tau$ is a tuple $(P, T, Pre, Post, m_0, I_s, \prec, \Lambda)$ where:

- $(P, T, Pre, Post, m_0)$ is a Petri Net where $P$ is a finite set of places, $T$ is a finite set of transitions with $P \cap T = \phi$, $m_0: P \to \mathbb{N}^+$ is the initial marking and $Pre, Post: T \to P \to \mathbb{N}$ are respectively the precondition and post-condition functions. For $f, g \in P \to \mathbb{N}^+, f \geq g$ means that $(\forall p \in P)(f(p) \geq g(p))$ and $f\{+,-\}g$ is $f(p)\{+,-\}g(p)$ for any $p$.

- $I_s: T \to \mathbb{I}^+$ is the static interval function which associates a firing temporal interval $I_s \in \mathbb{I}^+$ with each transition. The rational $\downarrow I_s(t)$ (resp. $\uparrow I_s(t)$) is the static earliest firing time (resp. the static latest firing time) of $t$ after the latter was enabled. Assuming that a transition $t$ became enabled at the last one at the time $\theta$, then $t$ can't be fired before $\theta + \downarrow I_s(t)$ and it must be done no later than $\theta + \uparrow I_s(t)$, unless disabled by firing some other transition. In this paper, intervals $[0,\infty[$ are omitted and **w** in the right end point of an interval denotes $\infty$. For example, $[3, w[$ denote the interval $[3, \infty[$.

- $\prec \subseteq T \times T$ is the priority relation, assumed irreflexive, asymmetric and transitive, between transitions. $t_1 \prec t_2$ means $t_2$ has priority over $t_1$.

- $\Lambda: T \to A_\tau$ is the labelling function that associates to each transition an operation.

The transitions of the net $\mathcal{M}$ (see section 3) are partitioned into purely transitions of the *SUT* model $\mathcal{M}_{SUT}$ (hence invisible for the environment $\mathcal{M}_E$ and labelled with $\tau$) and synchronizing transitions between the $\mathcal{M}_{SUT}$ and the ENV (hence observable for both parties). The set of transitions of the model $\mathcal{M}_{SUT}$ which are labelled with internal actions is $T_\tau = \{t \in T_{SUT} / \Lambda(t) = \tau\}$. The internal transitions are fired individually while synchronizing transitions are fired by complementary actions couples (e.g. $a?$ and $a!$). In a couple of synchronizing actions, we assume that the first component is an action of the SUT model $\mathcal{M}_{SST}$ while the second is of the environment model $\mathcal{M}_E$. A couple $(t, t') \in (T-T_\tau)^2$ is a synchronizing transition if $t$ and $t'$ are labeled with complementary synchronization actions which are noted $a, \bar{a}$ e.g. $\Lambda(t)=a?$ (resp. $a!$) and $\Lambda(t')=a!$ (resp. $a?$). We note $T_{SUT}$ the set of the *SUT* model transitions and $T_E$ the set of the environment model transitions. The set of the environment model transitions which complement a synchronizing transition $t \in T_{SUT}$ is equal to $CT_{SUT}(t) = \{t' \in T_E / \text{if } \Lambda(t) = a \text{ and } \Lambda(t') = \bar{a}\}$.

To illustrate the concepts, we use the light-controller model depicted by figure 3. The user interacts with the controller by touching a touch sensitive pad. The light has three intensity levels: OFF, DIMMED, and BRIGHT. Depending on the timing between successive touches, the controller toggles the light levels. For example, in DIM state, if a second touch is made quickly (before the switching time $T_{sw}=4$ time units) after the touch that caused the controller to enter dimmed state (from either OFF or BRIGHT state), the controller increases the level to bright. Conversely, if the second touch happens after the switching time, the controller switches the light OFF. If the light controller has been in OFF state for a long time (longer than or equal to Tidle = 20), it should reactivate upon a touch by going directly to bright level.

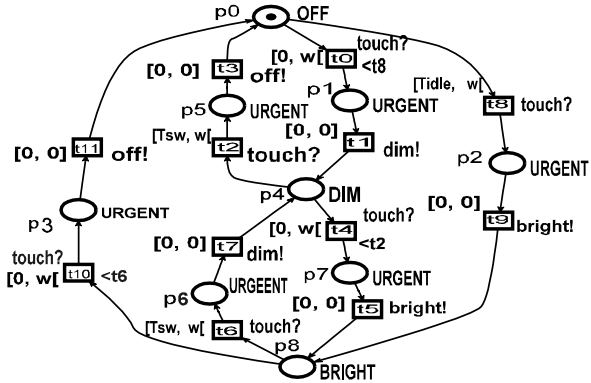

Figure 3. $\mathcal{M}_{SUT}$: the light controller model

The LPrTPN shown in Fig. 3 models a *SUT* which can be composed in parallel with the environment models shown in Fig. 4 & 5 respectively over the action $A_{in}=\{\text{touch}\}$ and $A_{out}=\{\text{off,dim,bright}\}$. We obtain two models ($\mathcal{M}_1=\mathcal{M}_{SUT}\|\mathcal{M}_{E1}$ and $\mathcal{M}_2=\mathcal{M}_{SUT}\|\mathcal{M}_{E2}$). Fig. 4 and 5 show two possible environment models, $\mathcal{M}_{E1}$ and $\mathcal{M}_{E2}$, for the simple light controller. Fig. 4 models a user capable of performing any sequence of touch actions. When the constant Treact is set to zero he is arbitrarily fast. A more realistic user is only capable of producing touches with a limited rate; this can be modelled setting Treact to a non-zero value. Fig 5 models a different user able to make two quick successive touches, but which then is required to pause for some time (to avoid cramp), *e.g.* Tpause = 5.

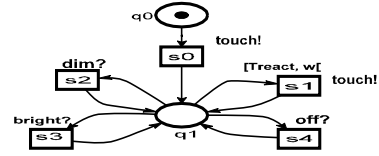

Figure 4. $\mathcal{M}_{E1}$ - a light switch controller environment model

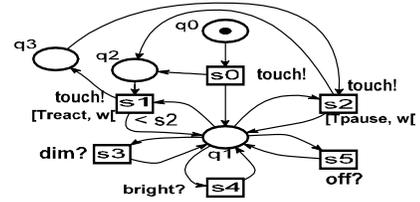

Figure 5. $\mathcal{M}_{E2}$ – another light switch controller environment model

### 4.1.3 Semantics

#### 4.1.3.1 Timed transition systems

Timed Transition Systems describe systems that combine discrete and continuous evolutions. They are used to define the behavior of timed systems such as LPrTPN. A Timed Transition System (TTS) is a transition system $S = (E, e_0, A_{S\tau}, \rightarrow)$, where $E$ is the set of states of the system, $e_0$ is the initial state, $A_{S\tau}$ is the set of actions (internal and couples of synchronizing actions). The transition relation $\rightarrow \subseteq E \times (A_{S\tau} \cup \mathbb{R}_{\geq 0}) \times E$ consists of discrete transitions $e \xrightarrow{(a,\bar{a})} e'$ or $e \xrightarrow{\tau} e'$ (with $(a,\bar{a}) \in A_S$) representing an instantaneous action, and continuous transitions $e \xrightarrow{d} e'$ (with $d \in \mathbb{R}_{\geq 0}$) representing the passage of $d$ units of time. Moreover, we require the following standard properties for *TTS*: (1) *Time-determinism*: if $e \xrightarrow{d} e'$ and $e \xrightarrow{d} e''$ with $d \in \mathbb{R}_{\geq 0}$ then $e'=e''$, (2) *0-delay*: $e \xrightarrow{0} e$, (3) *Additivity*: if $e \xrightarrow{d} e'$ and $e' \xrightarrow{d'} e''$, $(d, d' \in \mathbb{R}_{\geq 0})$ then $e \xrightarrow{d+d'} e''$, (4) *Continuity*: if $e \xrightarrow{d} e'$ then $\forall d', d'' \in \mathbb{R}_{\geq 0}$ such that $d = d'+d''$ there exists $e''$ such that $e \xrightarrow{d'} e'' \xrightarrow{d''} e'$.

### 4.1.3.2 States of an LPrTPN

A state of an LPrTPN is a pair $e=(m,I)$, where $m$ is a marking of the net with $m(p)$ the number of tokens in place $p$. A transition $t$ is enabled at marking $m$ iff $m \geq \text{Pre}(t)$. We denote by $En(m)$ the set of transitions enabled at $m$. It is then equal to $En(m)=\{t \in T / m \geq \text{Pre}(t)\}$. The second component of the pair $(m,I)$ is a partial function over $En(m)$ called the interval function. It associates exactly a temporal interval in $I^+$ with every enabled transition i.e. $I: En(m) \rightarrow I^+$. Intuitively, $I(t)$ represents the firing interval of the enabled transition $t$ shifted towards the origin as time elapses, and truncated to not-negative times. Assuming that the amount of time that has elapsed since $t$ is enabled for the last one is $\theta$ then $I(t)=I_s(t) \dot{-} \theta$. An enabled transition $t$ is fireable if (1) it is immediately fireable ($0 \in I(t)$), (2) no other transition with higher priority is fireable at the same instant, (3) if $t$ is not an internal transition then its complementary transition is also fireable. After the firing, some transitions are associated with their intervals $I_s(t)$ and we say that they are newly enabled.

The initial state of the LPrTPN $\mathcal{M}_I = \mathcal{M}_{SUT} || \mathcal{M}_{EI}$ that models the SUT "light-controller" with its associated environment (figures 3 and 4) is $e_0=(m_0, I_0)$, where:

- $m_0 = p_0(1), q_0(1)$ (places $p_0$ and $q_0$ are both marked with one token).
- $En(m_0)=\{t_0, t_8, s_0\}$ (transitions enabled by the initial marking $m_0$).
- $I_0(t_0)=[0,\infty[, I_0(t_8)=[\text{Tidle},\infty[$ and $I_0(s_0)=[0,\infty[$ (the interval function $I_0$ is $I_s$ restricted to the enabled transitions $En(m_0)$).
- Despite $I(t_0)=[0,\infty[$, $t_0$ is only fireable on $[0,\text{Tidle}[$ because $t_0 < t_8$ ($t_8$ has priority over $t_0$) and $I_0(t_8)=[\text{Tidle},\infty[$. By contrast, $t_8$ is fireable on $[\text{Tidle},\infty[$. The couple of transitions $(t_0,s_0)$ or $(t_8,s_0)$ labeled respectively by (touch?, touch!) can be fired respectively on $[0,\text{Tidle}[$ or on $[\text{Tidle},\infty[$.

The temporal information in states will be seen as firing domains instead of interval functions. The firing domain of a state $e=(m,I)$ is then described by an equations linear system with one variable per enabled transition (noted as transitions). The state will be then noted $e=(m,D)$ where $D=\{\underline{\phi}/(\forall t \in En(m))(\underline{\phi}_t = I(t))\}$. The state $e_0=(m_0,D_0)$ of the LPrTPN $\mathcal{M}_I = \mathcal{M}_{SUT} || \mathcal{M}_{EI}$ is:

$$e_0=(m_0,D_0) \text{ where } D_0: \begin{cases} 0 \leq t_0 \\ \text{Tidle} \leq t_8 \\ 0 \leq s_0 \end{cases}$$

### 4.1.3.3 Newly enabled Transition

For $m \in \mathbb{N}^+$ and $l \in T, t \in T_\tau$ such that $t \in En(m)$ we define a predicate $ne_\tau(l,m,t)$ which is true if $l$ is newly enabled by the firing of $t$ from $m$, and false otherwise. Formally, the predicate is defined by:

$$ne_\tau(l,m,t) = \begin{Bmatrix} l \in En(m-\text{Pre}(t)+\text{Post}(t)) \\ \wedge\ l \notin En(m-\text{Pre}(t)) \vee l=t \end{Bmatrix}$$

For $m \in \mathbb{N}^+, k \in T, t \in T_{SUT}$ and $t' \in CT_{SUT}(t)$ such that $t, t' \in En(m)$ we define a predicate $ne_{a,\bar{a}}(k,m,(t,t'))$, which is true if $k$ is newly enabled by the firing of $t$ and $t'$ simultaneously from marking $m$, and false otherwise by:

$$ne_{a,\bar{a}}(k,m,(t,t')) =$$
$$\begin{Bmatrix} k \in En(m-\text{Pre}(t)-\text{Pre}(t')+\text{Post}(t)+\text{Post}(t')) \\ \wedge\ k \notin En(m-\text{Pre}(t)-\text{Pre}(t')) \vee k=t \vee k=t' \end{Bmatrix}$$

The predicate $ne_{a,\bar{a}}(k,m,(t,t'))$ (resp. $ne_\tau(k,m,l)$) indicates the necessity to associate to $k$ its static interval after firing simultaneously the couple $(t,t')$ (resp. individually the transition $l$) at the marking $m$. Intuitively, it associates to the couple $(t,t')$ (resp. $l$) and to the transitions that could not be fired in parallel with $(t,t')$ (resp. $l$) their static intervals.

### 4.1.3.4 The semantics of an LPrTPN

The semantics of an $LPrTPN$ $\mathcal{N} = (P,T,\text{Pre},\text{Post},m_0,I_s,\prec,\Lambda)$ is a $TTS$ $[\![\mathcal{N}]\!] = (E,e_0,A_{S\tau},\rightarrow)$ where $E$ is the set of states $(m,I)$ of $\mathcal{N}$, $e_0$ its initial state and $\rightarrow \subseteq E \times (A_{S\tau} \cup \mathbb{R}_{\geq 0}) \times E$ consists of two kinds of transitions between states: discrete and continuous transitions. Discrete transitions are labeled with synchronizing or internal actions and continuous (or temporal) transitions are labeled by real values.

### 4.1.3.5 Transitions firing Algorithms

- The continuous transition relation is the result of time elapsing. It is defined by $(m,I) \xrightarrow{d} (m,I')$ iff
1. $d \in \mathbb{R}_{\geq 0}$
2. $(\forall t \in T)(t \in En(m) \Rightarrow d \leq \uparrow I(t))$

3. $(\forall t \in T)(t \in En(m) \Rightarrow I'(t) = I(t) \dot{-} \theta)$

A continuous transition of size $d$ is possible iff $d$ is not greater than the latest firing time of all enabled transitions (2). All firing intervals of enabled transitions are shifted synchronously towards the origin as time elapses, and truncated to non negative times (3).

- The discrete transitions are the result of the transitions firings of the Petri net. As it is showed above, they may be partitioned into internal independent and synchronizing transitions.

  ➤ the internal independent transition relation is defined by $(m, I) \xrightarrow{\tau} (m', I')$ iff
   1. $(\exists t \in T_{SUT})(t \in T_\tau \wedge t \in En(m))$
   2. $0 \in I(t)$
   3. $(\forall k \in T_{SUT})(k \in En(m) \wedge t \prec k \Rightarrow 0 \notin I(k))$
   4. $m' = m - Pre(t) + Post(t)$
   5. $(\forall k \in T)(m' \geq Pre(k) \Rightarrow I'(k) =$
      iff $ne_\tau(k,m,t)$ then $I_s(k)$ else $I(k)$

An internal transition $t$ of the $SUT$ model $\mathcal{M}_{SUT}$ may fire from a state $(m, I)$ if it is enabled at $m$ (1), immediately firable (2) and no internal or synchronizing transition of the $SUT$ model with higher priority satisfies these conditions (3). In the target state, the transitions of the combining model $\mathcal{M}$ that remained enabled while $t$ fired ($t$ excluded) retain their intervals, the others which are newly enabled by the result marking are associated with their static intervals (5).

  ➤ the synchronizing transition relation is defined by $(m, I) \xrightarrow{a, \bar{a}} (m', I')$ iff
   1. $(\exists (t, t') \in (T - T_\tau)^2)(\Lambda(t) = a \wedge \Lambda(t') = \bar{a} \Rightarrow t, t' \in En(m))$
   2. $0 \in I(t) \wedge 0 \in I(t')$
   2. $(\forall k \in T_{SUT})(k \in En(m) \wedge t \prec k \Rightarrow 0 \notin I(k))$
   3. $(\forall k' \in T_E)(k' \in En(m) \wedge t' \prec k' \Rightarrow 0 \notin I(k'))$
   4. $m' = m - (Pre(t) + Pre(t')) + Post(t) + Post(t')$
   6. $(\forall k \in T)(m' \geq Pre(k) \Rightarrow I'(k) =$
      iff $ne_{a,\bar{a}}(k, m, (t, t'))$ then $I_s(k)$ else $I(k)$

The synchronizing transitions $t$ and $t'$ labelled respectively by the complementary actions $a$ and $\bar{a}$ may fire simultaneously from the state $(m, I)$ if they are enabled (1), immediately fireable (2) and neither a transition of the $SUT$ model $\mathcal{M}_{SST}$ (a transition of $T_{SUT}$) nor a transition of the environment model $\mathcal{M}_E$ (a transition of $T_E$) with higher priorities compared to $t$ and $t'$ respectively satisfies these conditions (3&4). In the target state, the transitions that remained enabled while $t, t'$ fired ($t, t'$ being excluded) retain their intervals, the others which are newly enabled at the result marking are associated with their static intervals (6).

If the light controller and its environment (Fig. 3 and 4) remain in their initial state for 0.9 time units -the light controller doesn't receive any touch from the user- we have then a transition $e_0 \xrightarrow{0.9} e_1$. The new state $e_1 = (m_0, D_1)$ will be:

$m_0$: $p_0(1), q_0(1)$ and $D_1$: $\begin{cases} 0 \leq t_0 \\ \text{Tidle} - 0.9 \leq t_8 \\ 0 \leq s_0 \end{cases}$

The firing of the synchronizing transitions $(t_0, s_0)$ from $e_1$ leads to $e_2$ $\left(e_1 \xrightarrow{touch?, touch!} e_2\right)$. The state $e_2 = (m_1, D_2)$ is:

$m_1$: $p_1, q_1$ and $D_2$: $\begin{cases} 0 \leq t_1 \leq 0 \\ \text{Treact} \leq s_1 \leq \infty \\ 0 \leq s_2 \leq \infty \\ 0 \leq s_3 \leq \infty \\ 0 \leq s_4 \leq \infty \end{cases}$

The firing of $(t_1, s_1)$ from the state $e_2 = (m_1, D_2)$ leads to the state $e_3 = (m_2, D_3)$ $\left(e_2 \xrightarrow{dim!, dim?} e_3\right)$:

$m_2$: $p_4, q_1$ and $D_3$: $\begin{cases} \text{Tsw} \leq t_2 \leq \infty \\ 0 \leq t_4 \leq \infty \\ \text{Treact} \leq s_1 \leq \infty \\ 0 \leq s_2 \leq \infty \\ 0 \leq s_3 \leq \infty \\ 0 \leq s_4 \leq \infty \end{cases}$

With the properties of $TTS$, a *run* of $[\![\mathcal{N}]\!]$ can be defined as a finite sequence of moves $e_0 \xrightarrow{d_0} e'_0 \xrightarrow{\alpha_0} e_1 \xrightarrow{d_1} e'_1 \xrightarrow{\alpha_1} e_2 \cdots \xrightarrow{\alpha_n} e_{n+1}$ where discrete and continuous transitions alternate. $\alpha_{i, 0 \leq i \leq n}$ are either synchronizing transitions ($\alpha_i = (a, \bar{a}) \in A_s$) or pure transitions ($\alpha_i = \tau$) and $d_{i, 0 \leq i \leq n}$ are their relative firing times. To such a run corresponds the firing schedule which is the timed word $\omega = (\alpha_i, \eta_i)_{0 \leq i \leq n}$ over $A_{S\tau}$ where $\eta_i = \sum_{j=0}^{i} d_j$ is the firing time when the actions $(a, \bar{a})$ happen (resp. $\tau$ happens).

We denote by:

- *Support*($\omega$) the projection of the schedule $\omega$ over the alphabet $A_{S\tau}$. It is called its support. A schedule $\omega$ is realisable from a state $e$ if the discrete transitions of the support $\sigma = \alpha_0 \alpha_1 \cdots \alpha_n$ are successively fireable from $e$ at the associated firing times $\eta_0 \eta_1 \cdots \eta_n$.

- *Schedule*$_{ENV}(\omega)$ the projection of the schedule $\omega$ over the second components of the complementary actions, i.e. *Schedule*$_{ENV}(\omega) = (\bar{a}_i, \eta_i)_{0 \leq i \leq n}$ where $\bar{a}_i$ are the components of the environment model $\mathcal{M}_E$ and $\eta_i = \sum_{j=0}^{i} d_j$ is the firing time when the actions $(a_i, \bar{a}_i)$ happen. Note that the symbol $\tau$ doesn't appear in $Schedule_{ENV}(\omega)$, it is removed.

If the pausing time Tidle and the switching time Tsw in $\mathcal{M}_{SUT}$ are respectively equal 20 and 4 time units then:

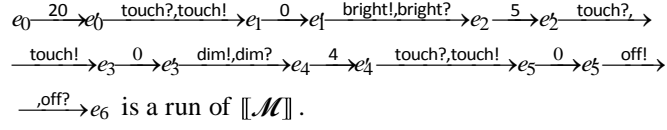

$e_0 \xrightarrow{20} e'_0 \xrightarrow{touch?,touch!} e_1 \xrightarrow{0} e'_1 \xrightarrow{bright!,bright?} e_2 \xrightarrow{5} e'_2 \xrightarrow{touch?,} \xrightarrow{touch!} e_3 \xrightarrow{0} e'_3 \xrightarrow{dim!,dim?} e_4 \xrightarrow{4} e'_4 \xrightarrow{touch?,touch!} e_5 \xrightarrow{0} e'_5 \xrightarrow{off!} \xrightarrow{,off?} e_6$ is a run of $[\![\mathcal{M}]\!]$.

$\omega = $ 20(touch?,touch!)20(bright!,bright?)25(touch?,touch!)25(dim!,dim?)29(touch?,touch!)29(off!,off?) is a firing schedule.

$Schedule_{ENV}(\omega) = $ 0.touch!.20.bright?.25.touch!.25.dim?.29.touh!.29.off?.

### 4.2 TINA (TIme Petri Net Analyzer)

TINA is a software environment for editing and analyzing (LPrT)PN [7]. It includes the tools:
– *nd (NetDraw):* is an editor for graphical or textual description of (*LPrT*)*PN*.
– *Tina:* For analysing (LPr)*TPN* models, it's necessary to finitely represent the state spaces by grouping some sets of states. *Tina* builds the *Strong State Classes Graph* (*SSCG*) proposed in [9], which preserves states and maximal traces of the state graph, and thus the truth value of all the formulae of the *SE-LTL* logic.
– *selt*: is a model checker for an enriched version of *State-Event LTL* [16], a *Linear Temporal Logic* supporting both *State* and *Event* properties. For the properties found false, *selt* produces a timed counter-example, namely the diagnostic (or witness) sequence. A diagnostic sequence of a property $\phi$ is a sequence of discrete transitions (complementary and/or internal transitions). A diagnostic trace is a schedule where its support is a diagnostic sequence. The firing of this schedule from $e_0$ allows satisfying the property $\phi$.
– *Plan:* is a path analysis tool. It computes all, or a single, timed firing sequence (schedule) over some given firing discrete transitions sequence. In particular, it computes the fastest schedules and shortest paths.

### 4.3 Deterministic Input Enabled and Output Urgent LPrTPN

To ensure time-optimal testability, the following semantic restrictions turn out to be sufficient. We define the notion of *Deterministic Input Enabled and Output Urgent LPrTPN*, *DIEOU-LPrTPN*, by restricting the underlying *TTS* defined by the *LPrTPN* as follows: (1) *Deterministic*: For every semantic state $e=(m,D)$ and $\gamma = A_{s\tau} \cup \mathbb{R}_{\geq 0}$, whenever $e \xrightarrow{\gamma} e'$ and $e \xrightarrow{\gamma} e''$ then $e'=e''$ ,(2) *(Weak) Input Enabled*: whenever $e \xrightarrow{d}$ for some delay $d \in \mathbb{R}_{\geq 0}$ then $\forall a \in A_{in} \; e \xrightarrow{a,\bar{a}}$ , (3) *Isolated Outputs*: $\forall a \in A_{out}$, whenever $e \xrightarrow{a,\bar{a}}$ and $e \xrightarrow{\beta}$ , $\beta \in A_{out} \times A_{in}$ then $a,\bar{a}=\beta$ , (4) *Output urgency*: whenever $e \xrightarrow{\alpha}$ , $\alpha \in A_{out} \cup \{\tau\}$ then $e \not\xrightarrow{d}, d \in \mathbb{R}_{\geq 0}$. These conditions are met by the model depicted by figure 3.

## 5. Test Generation

### 5.1 Conformance relation and test hypothesis

A conformance relation formalizes the set of *SUT* that behave correctly compared to a reference specification. In this paper, we require *Timed Input/Output Conformance relation* (*tioco*) [27] based on timed trace inclusion, *i.e.* the timed traces of the *SUT* are included in those of the specification. Thus after any input sequence, the *SUT* is allowed to produce an output only if the specification also able to produce that output. Similarly, the *SUT* may delay (staying silent) only if the specification also may delay.

A *SUT* is not a formal object (it is about a physical system or an implementation). However, formally proving its conformity requires modeling its semantics by a formal object. The remainder the paper assumes it can be modeled by an unknown *LPrTPN*. We assume that the tester can take the place of the environment and control the *SUT* via a distinguished set of observable input and output actions. For the *SUT* to be testable the *LPrTPN* of its specification should be controllable in the sense that it should be possible for an environment to drive the model through all of its syntactical parts (transitions and places). We therefore assume that the *SUT* specification is a *DIEOU-LPrTPN*, and that the *SUT* can be modelled by some unknown *DIEOU-LPrTPN*. The environment model need not be a *DIEOU-LPrTPN*. These assumptions are commonly referred to as the testing hypothesis.

To clarify the construction we may model the test case itself as an *LPrTPN* $\mathcal{M}_\lambda$ for the test sequence $\lambda$. Places in $\mathcal{M}_\lambda$ are labeled using two distinguished labels, **Pass** and **Fai**l. The execution of a test case is formalized as a parallel composition of the test case Petri net $\mathcal{M}_\lambda$ and the *SUT* model $\mathcal{M}_{SUT}$.

$$SUT \text{ passes } \mathcal{M}_\lambda \text{ iff } \mathcal{M}_\lambda \parallel \mathcal{M}_{SUT} \not\rightarrow \text{Fail}$$

$\mathcal{M}_\lambda$ is constructed such that a complete execution terminates in a **Fail** state (the place FAIL will be marked) if the *SUT* cannot perform $\lambda$ and such that it terminates in a **Pass** state (the place PASS will be marked) if the *SUT* can execute all actions of $\lambda$. The construction is illustrated in Fig. 6.

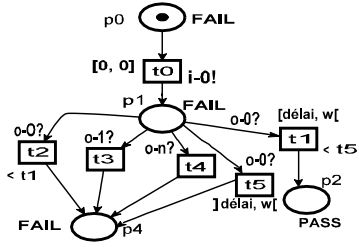

Figure 6. Test case LPrTPN $\mathcal{M}_\lambda$ for the sequence $\lambda = i_0!.delai.o_0?$

## 5.2 Test cases from *SE−LTL* properties

Let $\mathcal{M}$ be the *LPrTPN* of the *SUT* model together with its intended environment ENV; and $\phi$ the property, formulated in *SE−LTL*, to be verified over $\mathcal{M}$. As *SE−LTL* evaluate the properties on all possible executions, we consider the negation of $\phi$ *i.e.* the formula $\neg\phi$, then we submit it to *selt*. If the response of *selt* is negative, *i.e.* all the executions don't satisfy $\neg\phi$, so at least one satisfy its negation $\phi$. *selt* provide simultaneously a counter-example for $\neg\phi$, *i.e.* a diagnostic sequence $\sigma$ that demonstrates that property $\phi$ is satisfied. This sequence is submitted to the tool *plan* for computing a firing schedule $\omega$, or all the firing schedules, having this sequence as support. As we have seen in 3.2.1, $\omega$ is an alternating sequence of discrete transitions, synchronization (or internal) actions, performed by the system and its environment, and temporal constraints needed to reach the goal (the desirable state or event). Once $\omega$ is obtained, it is convenient to construct the associated test sequences. For *DIEOU-LPrTPN*, a test sequence is an alternating sequence of concrete delay actions and observable actions. Then a test sequence $\lambda$ is simply $Schedule_{ENV}(\omega)$. Finally, a test case to be executed on the real *SUT* implementation may be obtained from $\lambda$ by the addition of verdicts. Adding the verdicts depends on the chosen conformance relation between the specification and the *SUT*. The construction is illustrated in section 5.1. The test sequences produced by this technique are derived from the diagnostic traces, and are thus guaranteed to be included in the specification.

## 5.3 Test generation from a single test purpose

A test purpose is a property that the tester wants to observe on the SUT. A common approach to the generation of test cases is to first manually formulate a set of informal test purposes and then to formalize these such that the model can be used to generate one or more test cases for each test purpose. Because we use the diagnostic trace facility of the model-checker *selt*, the test purpose must be formulated as a *SE-LTL* property that can be checked by reachability analysis of the combined model $\mathcal{M}$. The test purpose can be directly transformed into a simple state or event reachability check. Also, the environment model can be replaced by a more restricted one that matches the behaviour of the test purpose only.

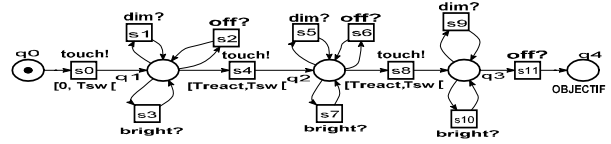

Figure 7. $\mathcal{M}_{E3}$, test environment for TP2

**TP1:** check that the light can become bright.
**TP2:** check that the light switches off after three successive touches.
**TP1** can be formulated as a simple *SE-LTL* state property $\phi_1 = \Diamond$BRIGHT or an event property $\phi_2 = \Diamond$bright! (eventually in some future the place BRIGHT of the light controller Petri net will be marked or the event bright! will be executed).
Among all diagnostic sequences exhibited by *selt* that satisfy the property $\phi_1$ (or $\phi_2$), two sequences are more interesting: the shortest and the fastest sequences. The two schedules associated to these sequences will be transformed to test cases as explained in 5.2.
For **TP1** we have:
- the shortest diagnostic sequence is
  (touch?,touch!)(bright!,bright?).
- The associated fastest schedule is :
  20.(touch?,touch!).20.(bright!,bright?)
- The test sequence is: 20.touch!.20.bright?
- The fastest sequence satisfying $\phi_1$ is: 0.(touch?,touch!) 0.(dim!,dim?).0.(touch?,touch!).0.(bright!,bright?)
- The test sequence is: 0.touch!.0.dim?.0.touch!.0.briht?

**TP2** can be formalized using the property $\mathcal{M}_{E3} \models \phi_3 = \Diamond$OBJECTIF with $\mathcal{M}_{E3}$ is the restricted environment model in Fig. 7. The fastest test sequence is 0.touch!.0.dim?.0.touch!.0.bright?. 0.touch!.0.off?.

## 5.4 Test Generation Based on Coverage criteria

A recurrent problem is to create a test suite that ensures that the specification or implementation is covered in some way. This ensures a certain level of systematicity is achieved in the test generation process. A large suite of coverage criteria may be proposed for the *LPrTPN* model, such as statements, transitions, places, markings and classes, each with its merits and application domain. In this paper, we use the following coverage criteria of the SUT model.

*Transition Coverage.* A test sequence satisfies the transition-coverage criterion if, when executed on the model, it fires every transition of the net. Transition coverage can be formulated by the property $\phi_t = \wedge_{i=1}^{N} \Diamond t_i$, where *N* is the number of transitions.

*Statement Coverage.* A test sequence satisfies the statement-coverage criterion if, when executed on the model, it executes all the observed actions. Statement coverage can be formulated by the property $\phi_s = \wedge_{i=1}^{N} \Diamond \Lambda(t_i)$ (in our example it is $\Diamond touch? \wedge \Diamond bright! \wedge \Diamond dim! \wedge \Diamond off!$).

*Place Coverage.* A test sequence satisfies the place-coverage criterion if, when executed on the model, it marks all the places of the net. Place coverage can be formulated by the property $\phi_m = \wedge_{i=1}^{M} \Diamond m(p_i)$, where *M* is the number of places of the net.

*Class Coverage*. A test sequence satisfies the class-coverage criterion if, when executed on the model, it generates the graph SSCG. We must first analyze the model with *Tina* and compute the *SSCG*. Second, we select a path in the *SSCG* graph that traverses all his nodes then compute particulars schedules with *plan*.

*Marking Coverage*. A test sequence satisfies the marking-coverage criterion if, when executed on the model, it generates the set $RM(\mathcal{M}_{SUT})$ of reachable markings. For generating test sequences that ensure this criterion, we compute the set $RM(\mathcal{M}_{SUT})$ by projecting SSCG over markings and finally encode the property in the *SE-LTL* logic.

In the example of the light controller, when the environment can touch arbitrarily, the generated fastest transition, statement, places covering test respectively are:
- **TC**:0.touch!.0.dim?.0.touch!.0.bright?.0.touch!.0.off?. 20.touch.20.bright?.24. touch.24. dim?.28.touch!.28.off? with an accumulated time of 28 t. u..
- **PC**: 20.touch.20.bright?.20.touch.20.off?.20.touch!. 20.dim?.20.touch!.20.bright?.24.touch!.24.dim!28.touch! . 28.off? with an accumulated time of 28 t. u..
- **CS**: 0.touch!.0.dim?.0.touch!.0.bright?.0.touch!.0. off? with 0 t. u..

## 5.5 Test Suite Generation

Frequently, for a given test purpose, we cannot obtain a single covering test sequence. This is due to the dead-ends in the model. To solve this problem, we allow for the model (and *SUT*) to be reset to its initial state and to continue the test after the reset to cover the remaining parts. The generated test will then be interpreted as a test suite consisting of a set of test sequences separated by resets (assumed to be implemented correctly in the *SUT*). To introduce resets in the model, we shall allow the user to designate some markings as being reset-able *i.e.* markings that allows to reach the initial marking $m_0$. Evidently, performing a reset may take some time Tr that must be taken into account when generating time optimal test sequences. Reset-able markings can be encoded into the model by adding reset transitions leading back to $m_0$. Let $m_r$ he reset-able marking, two reset transitions and a new place q must be added as:
The transition reset! must be added such as their input places are the encoded places (those of $m_r$) and its output place is q. The firing of reset! marks the place q. $(m_r,-) \xrightarrow{reset!} (q,[Tr,Tr]) \xrightarrow{\tau} (m_0, I_0)$.

## 5.6 Environment Behavior

Test sequences generated by the techniques presented above may be non-realizable; they may require the *SUT* environment to operate infinitely fast. We demonstrate how different environment assumptions influence the generated test sequences. Consider an environment where the user takes at least 2 time units between each touch action, such an environment can be obtained by setting the constant Treact to 2 in Fig. 4. The fastest test sequences become:
**TP1:** 0.touch!.0.dim?.2.touch!.2.bright?
**TP2:** 0.touch!.0.dim?.2.touch!.2.bright?.4.touch!.4.off?
Also re-examine the test suite **TC** generated by transition coverage, and compare with the one of execution time 32 generated when Treact equals 2.
**TC':**0.touch!.0.dim?.4.touch!.4.off?.24touch!.24.bright? .28.touch!.28.dim?30.touch!.30.bright?.32.touch!.32.off?
When the environment is changed to the pausing user (can perform 2 successive quick touches after which he is required to pause for some time: reaction time 2, pausing time 6), the fastest sequence has execution time 34, and follows a completely different strategy.
**TC'':**0.touch!.0.dim?.2.touch!.2.bright?.8.touch!.8.dim? .12.touch!.12.off?.32.touch!.32.bright?.34.touch!.34.off?

## 6. Conclusions

The paper proposes a method to transform the problem of timed test case generation from the LPrTPN model to a model-checking problem. Time-optimal test suites, which are computed from either a single test purpose or coverage criteria, may be generated using the *TINA* toolbox. Specifically, we used the tool *plan* to calculate the fastest and the shortest schedules associated with a diagnostic sequence issued by *selt* to derive test cases with optimal execution time. The transitions firings algorithms are revisited to the reactive character of real-time systems into account. The *DIEOU-LPrTPN* is quite restrictive, and generalization will benefit many real-time systems.